\documentclass[conference]{IEEEtran}
\IEEEoverridecommandlockouts
\usepackage{cite}
\usepackage{amsmath,amssymb,amsfonts}
\usepackage{amssymb}
\usepackage{algpseudocode}
\usepackage{amsfonts}
\usepackage{graphicx}
\usepackage{fancyhdr}  %
\usepackage{cases}
\usepackage{textcomp}
\usepackage{extarrows}
\usepackage{algorithm,algpseudocode}
\usepackage{multirow,tabularx}
\usepackage{mathtools}
\usepackage{xcolor}
\usepackage[english]{babel}
\usepackage{amsthm}

\def\BibTeX{{\rm B\kern-.05em{\sc i\kern-.025em b}\kern-.08em
    T\kern-.1667em\lower.7ex\hbox{E}\kern-.125emX}}

\newenvironment{shrinkeq}[1]
{ \bgroup
	\addtolength\abovedisplayshortskip{#1}
	\addtolength\abovedisplayskip{#1}
	\addtolength\belowdisplayshortskip{#1}
	\addtolength\belowdisplayskip{#1}}
{\egroup\ignorespacesafterend}

\begin{document}

%\title{Iterative Channel Estimation for RIS-aided Wireless Communication Systems
\title{Joint Channel Estimation and Signal Recovery in RIS-Assisted Multi-User MISO Communications}

%\author{Li Wei, Chongwen Huang, George~C.~Alexandropoulos, Zhaohui Yang, Chau~Yuen, and
%Zhaoyang Zhang 

%\thanks{Part of this work has been presented in \textit{IEEE SAM}, Hangzhou, China, 8--11 June 2020 \cite{parafac_SAM2020}.}
%\thanks{L. Wei, C. Huang, and C. Yuen are with the Engineering Product Development (EPD) Pillar, Singapore University of Technology and Design, Singapore 487372 (e-mails: wei\_li@mymail.sutd.edu.sg, \{chongwen\_huang, yuenchau\}@sutd.edu.sg).}
%
%\thanks{G.~C.~Alexandropoulos is with the Department of Informatics and Telecommunications, National and Kapodistrian University of Athens, Panepistimiopolis Ilissia, 15784 Athens, Greece (e-mail: alexandg@di.uoa.gr).}
%
%\thanks{Z. Zhang is with the Institute of Information and Communication
%Engineering, Zhejiang University, Hangzhou 310027, China
%(e-mail: ning\_ming@zju.edu.cn). }
%
%\thanks{M.~Debbah is with CentraleSup\'elec, University Paris-Saclay, 91192 Gif-sur-Yvette, France. M. Debbah is also with the Mathematical and Algorithmic Sciences Lab, Paris Research Center, Huawei Technologies France SASU, 92100 Boulogne-Billancourt, France (emails: merouane.debbah@huawei.com).

%}
	\author{
	\IEEEauthorblockN{Li Wei\IEEEauthorrefmark{1}, Chongwen Huang\IEEEauthorrefmark{2}, George~C.~Alexandropoulos\IEEEauthorrefmark{3},  Zhaohui Yang\IEEEauthorrefmark{4},  Chau~Yuen\IEEEauthorrefmark{1}, and Zhaoyang Zhang\IEEEauthorrefmark{2} }
	\IEEEauthorblockA{\IEEEauthorrefmark{1}Singapore University of Technology and Design, 487372 Singapore}
	\IEEEauthorblockA{\IEEEauthorrefmark{2}Department of Information and Electronic Engineering, Zhejiang University }
	\IEEEauthorblockA{\IEEEauthorrefmark{3}Department of Informatics and Telecommunications, National and Kapodistrian University of Athens, Greece}
	\IEEEauthorblockA{\IEEEauthorrefmark{4}Centre for Telecommunications Research, Department of Engineering, King's
		College London, WC2R 2LS, UK}
}

\maketitle

\begin{abstract}
Reconfigurable Intelligent Surfaces (RISs) have been recently considered as an energy-efficient solution for future wireless networks. Their dynamic and low-power configuration enables coverage extension, massive connectivity, and low-latency communications. Channel estimation and signal recovery in RIS-based systems are among the most critical technical challenges, due to the large number of unknown variables referring to the RIS unit elements and the transmitted signals. In this paper, we focus on the downlink of a RIS-assisted multi-user Multiple Input Single Output (MISO) communication system and present a joint channel estimation and signal recovery scheme based on the PARAllel FACtor (PARAFAC) decomposition. This decomposition unfolds the cascaded channel model and facilitates signal recovery using the Bilinear Generalized Approximate Message Passing (BiG-AMP) algorithm. The proposed method includes an alternating least squares algorithm to iteratively estimate the equivalent matrix, which consists of the transmitted signals and the channels between the base station and RIS, as well as the channels between the RIS and the multiple users. Our selective simulation results show that the proposed scheme outperforms a benchmark scheme that uses genie-aided information knowledge. We also provide insights on the impact of different RIS parameter settings on the proposed scheme.
\end{abstract}


\section{Introduction}\label{sec:intro}
Reconfigurable Intelligent Surfaces (RISs) are lately considered as a candidate technology for beyond fifth Generation (5G) wireless communication due to their potentially significant benefits in low powered, energy-efficient, high-speed, massive-connectivity, and low-latency communications\cite{Akyildiz2018mag,hu2018beyond,huang2019reconfigurable,Marco2019,qingqing2019towards,hcwjasc2020,alexandropoulos2020reconfigurable,9136592,9149091}. In its general form, a RIS is a reconfigurable planar metasurface composed of a large number of hardware-efficient passive reflecting elements \cite{Akyildiz2018mag,tang2019wireless,huang2018achievable}. Each unit element can alter the phase of the incoming signal without requiring a dedicated power amplifier, as for example needed in conventional amplify-and forward relaying systems \cite{Akyildiz2018mag,huang2018achievable,huang2019reconfigurable}.

The energy efficiency potential of RIS in the downlink of outdoor multi-user Multiple Input Single Output (MISO) communications was analyzed in \cite{huang2019reconfigurable}, while \cite{husha_LIS2} focused on an indoor scenario to illustrate the potential of RIS-based indoor positioning. It was shown in \cite{huang2018energy,han2019} that the latter potential is large even in the case of RISs with finite resolution unit elements and statistical channel knowledge. Recently, a novel passive beamforming and information transfer technique was proposed in \cite{yan2019passive} to enhance primary communication, as well as a two-step approach at the receiver to retrieve the information from both the transmitter and RIS. RIS-assisted communication in the millimeter wave \cite{wang2019intelligent} and THz \cite{ning2019intelligent} bands was also lately investigated.  However, the joint channel estimation and signal recovery problem in RIS-assisted systems has not been studied before. Most of the existing research works in RIS-assisted multi-user communications only focus on channel estimation.

The recent works \cite{Alkhateeb2019} and \cite{huang2019spawc} presented compressive sensing and deep learning approaches for recovering the involved channels and designing the RIS phase matrix. However, the deep learning approaches require extensive training during offline training phases, and the compressive sensing framework assumes that RIS has some active elements, specifically, a fully digital or hybrid analog and digital architecture attached at a RIS portion. The latter architectures increase the RIS hardware complexity and power consumption. A low-power receive radio frequency chain for channel estimation was considered in \cite{qingqing2019towards, 9053976}, which also requires additional energy consumption compared to passive RISs. Very recently, \cite{wei2020channel} and \cite{9104260} adopted PARAllel FACtor (PARAFAC) decomposition to estimate all involved channels, however, the signal recovery probem was not considered. In addition, some generalized approximate message passing algorithms are proposed for joint estimation and signal recovery, such as bilinear generalized approximate message passing (BiG-AMP) \cite{6898015}, bilinear adaptive generalized vector approximate message passing \cite{8580585}, and generalized sparse Bayesian learning algorithm \cite{8357527}.  In \cite{7574287}, the framework of expectation propagation was employed to perform joint channel estimation and decoding in massive multiple-input multiple-output systems with orthogonal frequency division multiplexing.

In this paper, motivated by the PARAFAC decomposition based channel estimation method \cite{wei2020channel, 9104260}, we present an efficient joint channel estimation and signal recovery technique for the downlink of RIS-aided multi-user MISO systems. Based on the PARAFAC decomposition, the proposed channel estimation method includes an Alternating Least Squares (ALS) algorithm to iteratively estimate two unkown matrices \cite{sidiropoulos2000blind}\cite{kolda2009tensor}. With the estimated matrix, the signal and another unknown channel can be estimated simultaneously using  BiG-AMP  \cite{6898015}. In such way,  we can estimate all involved channels and recover the transmitted signals  simultaneously.   Representative simulation results validate the accuracy of the proposed scheme and its superiority over benchmark techniques.

%The remainder of this paper is organized as follows. In Section \ref{sec:format}, the considered system model is given with the problem formulation. Details of the joint channel estimation and signal recovery scheme are provided in Section \ref{sec:channel_est}. Section~\ref{sec:simulation} presents the estimation performance under some considered scenarios. Finally, concluding remarks are drawn in Section~\ref{sec:conclusion}.

\textit{Notation}: Fonts $a$, $\mathbf{a}$, and $\mathbf{A}$ represent scalars, vectors, and matrices, respectively. $\mathbf{A}^T$, $\mathbf{A}^H$, $\mathbf{A}^{-1}$, $\mathbf{A^\dag}$, and $\|\mathbf{A}\|_F$ denote transpose, Hermitian (conjugate transpose), inverse, pseudo-inverse, and Frobenius norm of $ \mathbf{A} $, respectively. $a_{mn}$ is the $(m,n)$-th entry of $\mathbf{A}$. $|\cdot|$ and $(\cdot)^*$ denote the modulus and conjugation, respectively, while $\circ$ represents the Khatri-Rao  matrix product. Finally, notation $diag(\mathbf{a})$ represents a diagonal matrix with the entries of $\mathbf{a}$ on its main diagonal.

\section{System and Signal Models}\label{sec:format}
In this section, we first describe the system and signal models for the considered downlink RIS-assisted multi-user MISO communication system, and then present the considered PARAFAC decomposition for the end-to-end wireless communication channel.

\subsection{System Model}\label{subsec:signal model}
\begin{figure}\vspace{-0mm}
	\begin{center}
		\centerline{\includegraphics[width=0.42\textwidth]{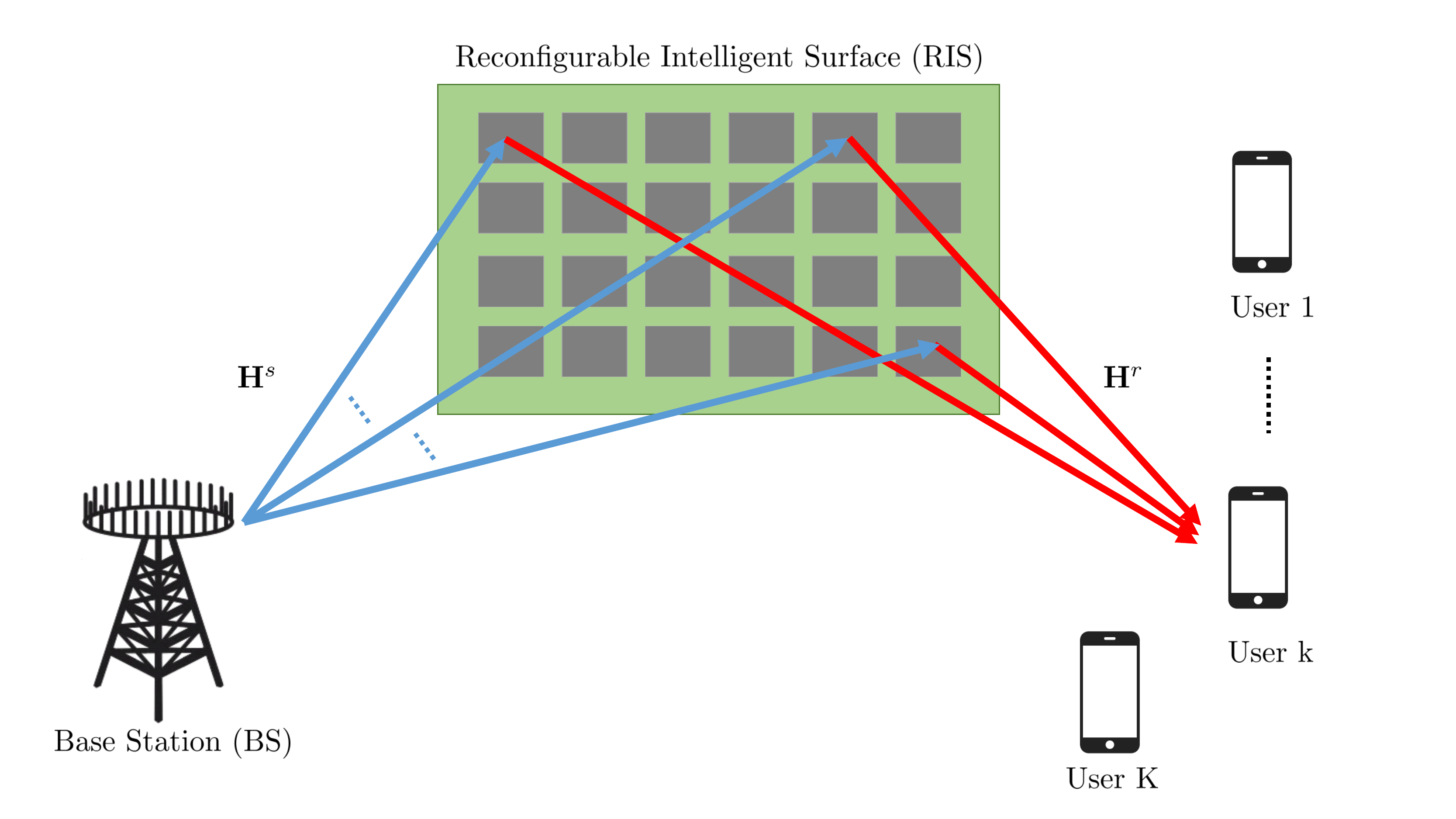}}  \vspace{-0mm}
		\caption{The considered RIS-based multi-user MISO system consisting of a $M$-antenna base station simultaneously serving in the downlink $K$ single-antenna mobile users.}
		\label{fig:Estimation_Scheme} \vspace{-4mm}
	\end{center}
\end{figure}

We consider the downlink communication between a Base Station (BS) equipped with $M$ antenna elements and $K$ single-antenna mobile users. We assume that this communication is realized via a discrete-element RIS deployed on the facade of a building existing in the vicinity of the BS side, as illustrated in Fig$.$~\ref{fig:Estimation_Scheme}. The RIS is comprised of $N$ unit cells of equal small size, each made from metamaterials capable of adjusting their reflection coefficients. The direct signal path between the BS and the mobile users is neglected due to unfavorable propagation conditions, such as large obstacles. The received discrete-time signals at all $K$ mobile users for $T$ consecutive time slots using the $p$-th RIS phase configuration (out of the $P$ available in total, hence, $p=1,2,\ldots,P$) can be compactly expressed with $\mathbf{Y}_{p}\in\mathbb{C}^{K \times T}$ given by
\begin{equation}\label{equ:YXZ}
\mathbf{Y}_{p}\triangleq\mathbf{H}^{r} D_p(\mathbf{\Phi})  \mathbf{H}^{s} \mathbf{X}+\mathbf{W}_{p},
\end{equation}
where $D_p(\mathbf{\Phi})\triangleq diag (\mathbf{\Phi}_{p,:})$ with $\mathbf{\Phi}_{p,:}$ representing the $p$-th row of the $P\times N$ complex-valued matrix $\mathbf{\Phi}$. Each row of $\mathbf{\Phi}$ includes the phase configurations for the RIS unit elements, which are usually chosen from low resolution discrete sets \cite{huang2018energy}. $\mathbf{H}^{s}\in\mathbb{C}^{N\times M}$ and $\mathbf{H}^{r}\in\mathbb{C}^{K\times N}$ denote, respectively, the channel matrices between RIS and BS, and between all users and RIS. Additionally, $\mathbf{X}\in\mathbb{C}^{M \times T}$ includes the BS transmitted signal within $T$ time slots (it must hold $T \geq M$ for channel estimation), and $\mathbf {W}_p \in \mathbb{C}^{K \times T}$ is the Additive White Gaussian Noise (AWGN) matrix having zero mean and unit variance elements.

\subsection{Decomposition of the Received Training Signal}\label{subsec:prob_formu}
The channel matrices $\mathbf{H}^{s}$ and $\mathbf{H}^{r}$ in \eqref{fig:Estimation_Scheme} are in general unknown and need to be estimated. We hereinafter assume that these matrices have independent and identically distributed complex Gaussian entries; the entries between the matrices are also assumed independent. Our objective in this work is to continuously track $\mathbf{H}^{s}$, $\mathbf{H}^{r}$, and $\mathbf{X}$ under the known observation signals $\textbf{Y}$ and few training pilots. To achieve this goal, we propose the following two-stage method. Firstly, we estimate the channel $\mathbf{H}^r$ and the equivalent matrix $\mathbf{H}^e\triangleq\mathbf{H}^s \mathbf{X} \in \mathbb{C}^{N \times T}$ using in \eqref{equ:YXZ} and the PARAFAC decomposition, according to which the intended channels are represented using tensors \cite{kroonenberg1980principal,rong2012channel}. Then, the channel $\mathbf{H}^s$ and the transmitted signal $\mathbf{X}$ are simultaneously estimated using the BiG-AMP algorithm \cite{6898015}.  Thus, each $(k,t)$-th entry of  $\mathbf{Z}_{p}$, which is the noiseless version of $\mathbf{Y}_p$, with $k=1,2,\ldots,K$ and $t=1,2,\ldots,T$ is obtained as
\begin{equation}\label{scalar_decomp}
[\mathbf{Z}_{p}]_{k t}=\sum_{n=1}^{N} [\mathbf{H}^{r}]_{k n} [\mathbf{H}^{e}]_{n t} [\mathbf{\Phi}]_{p n},
\end{equation}
where $[\mathbf{H}^{e}]_{n t}$, $[\mathbf{H}^{r}]_{k n}$, and $[\mathbf{\Phi}]_{p n}$ denote the $(n,t)$-th entry of $\mathbf{H}^{e}$, the $(k,n)$-th entry of $\mathbf{H}^{r}$, and the $(p,n)$-th entry of $\mathbf{\Phi}$, respectively, with $n=1,2,\ldots,N$.

Resorting to the PARAFAC decomposition \cite{harshman1994parafac,bro2003new,ten2002uniqueness,roemer2008closed}, each matrix $\mathbf{Z}_{p}$ in \eqref{scalar_decomp} out of the $P$ in total can be represented using three matrix forms. These matrices form the horizontal, lateral, and frontal slices of the tensor composed in \eqref{scalar_decomp}. The unfolded forms of the mode-1, mode-2, and mode-3 of $\mathbf{Z}_{p}$'s are given in \cite[eqs. (6)--(8)]{wei2020channel}. In the sequel, we present the proposed joint channel estimation and signal recovery scheme for the simultaneous estimation of the channel matrices $\mathbf{H}^{r}$ and $\mathbf{H}^{s}$, and the recovery of the transmitted signals $\mathbf{X}$.

%are expressed as follows:
%\begin{equation}\label{mode_1}
%\text{Mode-1:} \quad {\mathbf{Z}_1=({\mathbf{H}^e}^T \circ   \mathbf{\Phi}){\mathbf{H}^r}^T}\in\mathbb{C}^{PT \times K},
%\end{equation}
%\begin{equation}\label{mode_2}
%\text{Mode-2:}  \quad {\mathbf{Z}_2=( \mathbf{\Phi} \circ \mathbf{H}^r)\mathbf{H}^e}\in\mathbb{C}^{KP \times T},
%\end{equation}
%\begin{equation}\label{mode_3}
%\quad \text{Mode-3:} \quad {\mathbf{Z}_3=( \mathbf{H}^r \circ  {\mathbf{H}^e}^T)  \mathbf{\Phi}^T}\in\mathbb{C}^{TK \times P}.
%\end{equation}

%In the following section,  we present the joint estimation and recovery approach based on these unfolded forms.

\section{Propose Joint Channel Estimation\\ and Signal Recovery}\label{sec:channel_est}
\begin{figure*}\vspace{-1mm}
	\begin{center}
		\centerline{\includegraphics[width=1\textwidth]{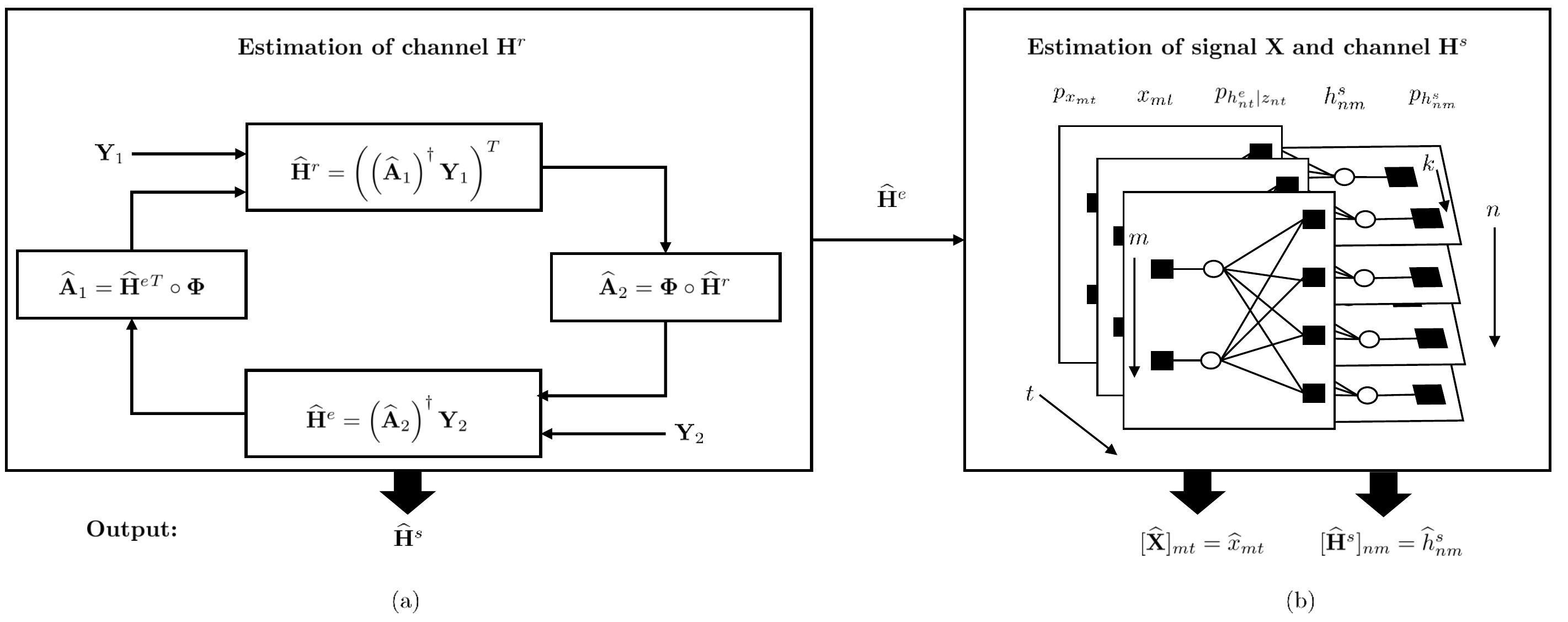} } \vspace{-2mm}
		\caption{The proposed framework for the joint channel estimation and signal recovery in RIS-assisted multi-user MISO systems.}
		\label{fig:framework}
      \vspace{-6mm}
	\end{center}
\end{figure*}

The proposed joint channel estimation and signal recovery scheme, which is graphically summarized in Fig$.$~\ref{fig:framework}, is composed of two main blocks. The first block deals with the estimation of the channels $\mathbf{H}^s$ and $\mathbf{H}^e$. Capitalizing on the latter estimation, the second block simultaneously estimates the channel $\mathbf{H}^r$ and recovers the transmitted signal $\mathbf{X}$. The details of the considered algorithms for these two blocks, as well as the treatment of their possible ambiguities, are presented in following subsections.

  \vspace{-2mm}

\subsection{The First Block}
In this block, the ALS-based channel estimation algorithm of \cite{wei2020channel} is adopted. Specifically, the channels $\mathbf{H}^r$ and $\mathbf{H}^e$ are obtained in an iterative manner by alternatively minimizing conditional Least Square (LS) criteria using the tensor ${\mathbf{Y}}$ for the received signal and the noiseless unfolded forms for all $P$ matrices $\mathbf{Z}_p$ (similar to \cite[eqs. (6)--(8)]{wei2020channel}). As shown in Fig$.$~\ref{fig:framework} (a), two intermediate matrices $\mathbf{A}_1={\mathbf{H}^e}^T\circ \mathbf{\Phi} \in \mathbb{C}^{PT \times N}$ and $\mathbf{A}_2= \mathbf{\Phi} \circ\mathbf{H}^r \in \mathbb{C}^{KP \times N}$ are iteratively updated based on the estimates for $\mathbf{H}^r$ and $\mathbf{H}^e$ at each iteration. At the $i$-th algorithmic iteration, the $i$-th estimation for $\mathbf{H}^r$, denoted by $\widehat{\mathbf{H}}^r_{(i)}$, is obtained from the minimization of the following LS objective function:
\begin{shrinkeq}{-1.5ex}
\begin{equation}\label{ls_hr}
\begin{aligned}
 J\left(\widehat{\mathbf{H}}^r_{(i)}\right)=\left\|{\mathbf{Y}_1}-\widehat{\mathbf{A}}_{1}^{(i-1)}\left(\widehat{\mathbf{H}}^r_{(i)}\right)^T\right\|_{F}^{2},
\end{aligned}
\end{equation}
\end{shrinkeq}
where $\widehat{\mathbf{A}}_{1}^{(i-1)}=\widehat{\mathbf{H}}^e_{(i-1)}\circ\mathbf{\Phi}$. The solution of \eqref{ls_hr} is given by
\begin{shrinkeq}{-1ex}
\begin{equation}\label{ls_hr_solution}
\left(\widehat{\mathbf{H}}^r_{(i)}\right)^T=\left(\widehat{\mathbf{A}}_{1}^{(i-1)}\right)^{\dagger}{\mathbf{Y}_1}.
\end{equation}
\end{shrinkeq}

Similarly, for the $\mathbf{H}^e$ estimation, we use the mode-2 unfolded form and formulate the following LS objective function for the $i$-th estimation for $\mathbf{H}^e$, which is denoted by $\widehat{\mathbf{H}}^e_{(i)}$:
\begin{shrinkeq}{-1ex}
\begin{equation}\label{ls_hs}
\begin{aligned} J\left(\widehat{\mathbf{H}}^e_{(i)}\right) &=\left\|\mathbf{Y}_2-\widehat{\mathbf{A}}_{2}^{(i)}\mathbf{H}^e \right\|_{F}^{2},
\end{aligned}
\end{equation}
\end{shrinkeq}
where ${\mathbf{Y}_2}\in\mathbb{C}^{KP \times T}$ is the mode-2 matrix-stacked form of tensor ${\mathbf{Y}}$, and $\widehat{\mathbf{A}}_{2}^{(i)}= \mathbf{\Phi}\circ\widehat{\mathbf{H}}^r_{(i)} $. The solution of \eqref{ls_hs} is easily obtained as
\begin{shrinkeq}{-1.5ex}
\begin{equation}\label{ls_hs_solution}
\widehat{\mathbf{H}}^e_{(i)}=\left(\widehat{\mathbf{A}}_{2}^{(i)}\right)^{\dagger} {\mathbf{Y}_2}.
\end{equation}
\end{shrinkeq}
The latter estimation method is summarized in Algorithm~\ref{alg:als}.

\setlength{\textfloatsep}{0.5cm}
\setlength{\floatsep}{0.5cm}
\begin{algorithm}[!t] \caption{\textbf{Estimation of Channel $\mathbf{H}^r$ and $\mathbf{H}^e$  \cite{wei2020channel}}  }
	\label{alg:als}
	\begin{algorithmic}[1]
		\Require A feasible $\mathbf{\Phi}$, $\epsilon>0$, and the number of maximum algorithmic iterations $I_{\rm max}$.
		\State \textbf{Initialization:} Initialize with a random feasible phase matrix $\mathbf{\Phi}$ and $\widehat{\mathbf{H}}^e_{(0)}$ obtained from the $N$ non-zero eigenvalues of $\mathbf{Y}_2^H \mathbf{Y}_2$, and set algorithmic iteration $i=1$.
		\For {$i=1,2,\ldots,I_{\rm max}$}
		\State Obtain $\widehat{\mathbf{H}}^r_{(i)}$ using $\widehat{\mathbf{H}}^r_{(i)}=\left(\left(\widehat{\mathbf{A}}_{1}^{(i-1)}\right)^{\dagger} {\mathbf{Y}_1}\right)^T$.
		\State Obtain $\widehat{\mathbf{H}}^e_{(i)}$ using $\widehat{\mathbf{H}}^e_{(i)}=\left(\widehat{\mathbf{A}}_{2}^{(i)}\right)^{\dagger} {\mathbf{Y}_2} $.
		\State \textbf{Until}   $\frac{\|\widehat{\mathbf{H}}^{r}_{(i)}-\widehat{\mathbf{H}}^{r}_{(i-1)} \|_{F}^{2}}
		{ \|\widehat{\mathbf{H}}^{r}_{(i)} \|_{F}^{2}} \leq \epsilon$, or $i=I_{\rm max}$.
		\EndFor
		\Ensure $\widehat{\mathbf{H}}^r_{(i)}$ and $\widehat{\mathbf{H}}^e_{(i)}$ that are the estimations of $\mathbf{H}^r$ and $\mathbf{H}^e$, respectively.
	\end{algorithmic}
\end{algorithm}

\subsection{The Second Block}
In this block, we jointly estimate $\mathbf{H}^s$ and $\mathbf{X}$ given the estimated matrix $\mathbf{H}^e=\mathbf{H}^s \mathbf{X}$ and using the BiG-AMP algorithm \cite{6898015}. First, we derive the following posterior distribution:
\begin{equation} \label{equ:post}
\begin{aligned}
& p_{\mathbf{X}, \mathbf{H}^s \mid \mathbf{H}^e}(\mathbf{X}, \mathbf{H}^s \mid \mathbf{H}^e) \\
=& p_{\mathbf{H}^e \mid \mathbf{X}, \mathbf{H}^s}(\mathbf{H}^e \mid \mathbf{X}, \mathbf{H}^s) p_{\mathbf{X}}(\mathbf{X}) p_{\mathbf{H}^s}(\mathbf{H}^s) / p_{\mathbf{H}^e}(\mathbf{H}^e) \\
\propto & p_{\mathbf{H}^e \mid \mathbf{Z}}(\mathbf{H}^e \mid \mathbf{H}^s \mathbf{X}) p_{\mathbf{X}}(\mathbf{X}) p_{\mathbf{H}^s}(\mathbf{H}^s) \\
=&\left[\prod_{n=1}^{N} \prod_{t=1}^{T} p_{h^e_{n t} \mid z_{n t}}\left(h^e_{n t} \mid \sum_{k=1}^{M} h^s_{n k} x_{k t}\right)\right] \\
& \times\left[\prod_{m=1}^{M} \prod_{t=1}^{T} p_{x_{m t}}\left(x_{m t}\right)\right]\left[\prod_{n=1}^{N} \prod_{m=1}^{M} p_{h^s_{n m}}\left(h^s_{n m}\right)\right]
\end{aligned}
\end{equation}
This joint distribution of $\mathbf{H}^s$ and $\mathbf{X}$ given $\mathbf{H}^e$ can be represented with the factor graph in Fig.~\ref{fig:framework} (b). In this graph, the factor nodes are represented with black boxes, while the variable nodes are represented with white circles. The estimated $\widehat{\mathbf{H}}^e$ from the first block is treated as the parameters of the $p_{\mathbf{H}^e \mid \mathbf{Z}}$ factor nodes. Applying the BiG-AMP algorithm, the channel $\mathbf{H}^s$ and signal $\mathbf{X}$ can be estimated, as shown in Algorithm~2.

In Algorithm~2, Steps $3$ and $4$ compute the `plug in' estimate of $\widehat{\mathbf{H}}^e$ and its corresponding variances. Based on this estimate, the Onsager correction is applied in Steps $5$ and $6$. Afterwards, Steps $7$ and $8$ calculate the marginal posterior means and variances, respectively, of $\widehat{\mathbf{H}}^e$. Considering that $\widehat{\mathbf{H}}^e$ is the output of the first stage (i.e., the output of Algorithm~1), we assume the noise to be zero, thus the scaled residual and corresponding variances are computed in Steps $9$ and $10$ without the AWGN term. In the sequel, steps $11$ and $12$ compute the means and variances of the corrupted version of the estimation for $\mathbf{X}$, and then, Steps $13$ and $14$ calculate the means and variances of the AWGN-corrupted observation of the estimated channel $\mathbf{H}^s$. Finally, based on the priors of $\mathbf{X}$ and $\mathbf{H}^s$, the posteriors of $\mathbf{X}$ and $\mathbf{H}^s$ are computed in Steps $15$ until $18$, as follows:
\begin{shrinkeq}{-1ex}
\begin{equation} \label{equ:pos_x}
\begin{array}{l}
p_{x_{m t} \mid {r}_{m t}}\left(x_{m t} \mid \widehat{r}_{m t}(i) ; \nu_{m t}^{r}(i)\right) \\
\propto  p_{x_{m t}}\left(x_{m t}\right) \mathcal{N}\left(x_{m t} ; \widehat{r}_{m t}(i), \nu_{m t}^{r}(i)\right)
\end{array}
\end{equation}
\end{shrinkeq}
and
\begin{shrinkeq}{-1ex}
\begin{equation}\label{equ:pos_h}
\begin{array}{l}
p_{{h}^s_{n m} \mid {q}_{nm}}\left(h^s_{n m} \mid \widehat{q}_{n m}(i) ; \nu_{n m}^{q}(i)\right) \\
\propto   p_{{h}^s_{n m}}\left(h^s_{n m}\right) \mathcal{N}\left(h^s_{n m} ; \widehat{q}_{n m}(i), \nu_{n m}^{q}(i)\right)
\end{array}
\end{equation}
\end{shrinkeq}

The posterior probability \eqref{equ:pos_x} is used to compute the variances and means of $x_{m t}$ $\forall$$m,t$ in Step $15$ and $16$. In fact, the probability \eqref{equ:pos_x} can be interpreted as the exact posterior probability of each $x_{m t}$ given the observation $r_{m t}$ under the prior model $p_{x_{m t}}\left(x_{m t}\right)$ and the likelihood model $p_{{r}_{m t} \mid {x}_{m t}}\left(\widehat{r}_{m t}(i) \mid x_{m t} ; \nu_{m t}^{r}(i)\right)=\mathcal{N}\left(\widehat{r}_{m t}(i) ; x_{m t}, \nu_{m t}^{r}(i)\right)$. The latter model is assumed in each $i$-th iteration of the BiG-AMP algorithm. Similarly, the posterior probability \eqref{equ:pos_h} is used to compute the variances and means of $h^s_{n m}$ in Steps $15$ and $16$. It can be interpreted as the posterior probability of $h^s_{n m}$ given the observation $q_{n m}$ under the prior model  $p_{h^s_{n m}}\left(h^s_{n m}\right)$ and the likelihood model $p_{{q}_{n m} \mid {h}^s_{n m}}\left(\widehat{q}_{n m}(i) \mid h^s_{n m} ; \nu_{n m}^{q}(i)\right)=\mathcal{N}\left(\widehat{q}_{m t}(i) ; h^s_{n m}, \nu_{n m}^{q}(i)\right)$.

\begin{algorithm}[!t] \caption{\textbf{The BiG-AMP Algorithm \cite{6898015}}}
	\label{alg:bigamp}
	\begin{algorithmic}[1]
		\Require $\widehat{\mathbf{H}}^e$, $\epsilon>0$, and the number of maximum algorithmic iterations $I_{\rm max}$.
		\State \textbf{Initialization:} Initialize $\hat{s}_{n t}(0)=0$, and choose $\nu_{m t}^{x}(1), \widehat{x}_{m t}(1), \nu_{n m}^{h}(1), \widehat{h}^s_{n m}(1), n=1,\ldots,N, t=1,\ldots,T, m=1,\ldots,M$.
		\For {$i=1,2,\ldots,I_{\rm max}$}
		\State  $\forall n,t: \bar{\nu}_{n t}^{p}(i)=\sum_{n=1}^{N}\left|\widehat{h}^s_{n m}(i)\right|^{2} \nu_{m t}^{x}(i)+\nu_{n m}^{h}(i)\left|\widehat{x}_{m t}(i)\right|^{2}$
		\State  $\forall n,t: \bar{p}_{n t}(i)=\sum_{n=1}^{N} \widehat{h}^s_{n m}(i) \widehat{x}_{m t}(i)$
		\State  $\forall n,t: \nu_{n t}^{p}(i)=\bar{\nu}_{n t}^{p}(i)+\sum_{n=1}^{N} \nu_{n m}^{h}(i) \nu_{m t}^{x}(i)$
		\State  $\widehat{p}_{n t}(i)=\bar{p}_{n t}(i)-\widehat{s}_{n t}(t-1) \bar{\nu}_{n t}^{p}(i)$
		\State  $\forall n,t: \nu_{n t}^{z}(i)=\operatorname{var}\left\{z_{n t} \mid \mathrm{p}_{n t}=\widehat{p}_{n t}(i) ; \nu_{n t}^{p}(i)\right\}$
		\State  $\forall n,t: \widehat{z}_{n t}(i)=\mathrm{E}\left\{\mathrm{z}_{n t} \mid \mathrm{p}_{n t}=\widehat{p}_{n t}(i) ; \nu_{n t}^{p}(i)\right\}$
		\State  $\forall n,t: \nu_{n t}^{s}(i)=\left(1-\nu_{n t}^{z}(i) / \nu_{n t}^{p}(i)\right) / \nu_{n t}^{p}(i)$
		\State  $\forall n,t: \widehat{s}_{n t}(i)=\left(\widehat{z}_{n t}(i)-\widehat{p}_{n t}(i)\right) / \nu_{n t}^{p}(i)$
		\State  $\forall m,t: \nu_{m t}^{r}(i)=\left(\sum_{m=1}^{M}\left|\widehat{h}^s_{n m}(i)\right|^{2} \nu_{n t}^{s}(i)\right)^{-1}$
		\State  $\begin{aligned}\forall m,t:  \widehat{r}_{m t}(i)&\!=\!\widehat{x}_{m t}(i)\!\left(\!1\!-\!\nu_{m t}^{r}(i) \!\sum_{m=1}^{M} \nu_{n m}^{h}(i) \nu_{n t}^{s}(i)\!\right) \\ &+\nu_{m t}^{r}(i) \sum_{m=1}^{M} \widehat{h}_{n m}^{s*}(i) \widehat{s}_{n t}(i) \end{aligned}$ 
		\State  $\forall n,m: \nu_{n m}^{q}(i)=\left(\sum_{l=1}^{L}\left|\widehat{x}_{m t}(i)\right|^{2} \nu_{n t}^{s}(i)\right)^{-1}$
		\State  $\begin{aligned}\forall n,m:  \widehat{q}_{n m}(i)& \!=\!\widehat{h}^s_{n m}(i)\! \left(\! 1 \!-\!\nu_{n m}^{q}(i)\! \sum_{l=1}^{L} \nu_{m t}^{x}(i)  \nu_{n t}^{s}(i)\!\right) \\ &+\nu_{n m}^{q}(i) \sum_{l=1}^{L} \widehat{x}_{m t}^{*}(i) \widehat{s}_{n t}(i) \end{aligned}$
		\State  $\nu_{m t}^{x}(i+1)=\operatorname{var}\left\{x_{m t} \mid r_{m t}=\widehat{r}_{m t}(i) ; \nu_{m t}^{r}(i)\right\}, \forall m,t$
		\State  $\widehat{x}_{m t}(i+1)=\mathrm{E}\left\{\mathrm{x}_{m t} \mid \mathrm{r}_{m t}=\widehat{r}_{m t}(i) ; \nu_{m t}^{r}(i)\right\}, \forall m,t$
		\State  $\nu_{n m}^{h}(i+1)=\operatorname{var}\left\{h^s_{n m} \mid q_{n m}=\widehat{q}_{n m}(i) ; \nu_{n m}^{q}(i)\right\}, \forall n,m$
		\State  $ \widehat{h}^s_{n m}(i+1)=\mathrm{E}\left\{{h}^s_{n m} \mid \mathrm{q}_{n m}=\widehat{q}_{n m}(i) ; \nu_{n m}^{q}(i)\right\}, \forall n,m$
	 
		\State \textbf{Until} $\sum_{m, l}\left|\bar{p}_{n t}(i)-\bar{p}_{n t}(i-1)\right|^{2} \!\leq \!\epsilon \sum_{m, l}\left|\bar{p}_{n t}(i)\right|^{2}$.
		\EndFor
		\Ensure $[ \widehat{\mathbf{H}}^s ]_{n m}=\widehat{h}^s_{n m}$ and $[\widehat{\mathbf{X}}]_{m t}=\widehat{x}_{m t}$, respectively.
	\end{algorithmic}
\end{algorithm}

\subsection{Ambiguity Removing}
The proposed joint estimation and signal recovery scheme involves ambiguities in the two blocks in Fig.~\ref{fig:framework}. In the first block, the iterative estimation of $\mathbf{H}^e$ and $\mathbf{H}^r$, using the ALS approach included in Algorithm~1, encounters a scaling ambiguity from the convergence point, which can be resolved with adequate normalization \cite{rong2012channel,lioliou2009performance}. To tackle this issue, the first column of $\mathbf{H}^r$ has been normalized.

In the second block, phase and scaling ambiguities exist in the estimation of $\mathbf{H}^s$ and $\mathbf{X}$ using Algorithm~2. To remove the latter ambiguity, we assume initial pilot signal transmission in $\mathbf{X}$. For the phase ambiguity, we adopt the method in \cite{8879620}. In particular, the transmitted signal is designed to be a full-rank matrix. It is generated from the Bernoulli Gaussian distribution with sparsity $\beta$ ($\beta$ is the portion of nonzero elements in transmitted signal), and sparsity pattern known to the receivers. With the latter settings, the ambiguities can be removed from the proposed scheme.

\subsection{Computational complexity}
In Algorithm~1, the computational complexities for the estimations of $\mathbf{H}^s$ and $\mathbf{H}^e$ are $\mathcal{O}(N^3+4N^2KP-NKP+2NMKP-NM)$ and $\mathcal{O}(N^3+4N^2MP-NMP+2NMKP-NK)$, respectively. Thus, using the feasibility condition in the proposed PARAFAC-based structure \cite{wei2020channel}, the total computational complexity for this algorithm becomes $\mathcal{O}(2N^3+4NMKP)$. The complexity of Algorithm~2 is dominated by the matrix multiplications per algorithmic iteration. In Steps $3$ up to $5$, as well as from Steps $11$ until $14$, each computation requires $NMT$ multiplications, where the computations within Steps $6$ and $10$ are of the order $\mathcal{O}(NM+MT)$. Thus, Algorithm~2 requires $\mathcal{O} (NMT+NM+MT)$ mathematical operations per iteration. Putting all above together, the total computational complexity of the proposed joint channel estimation and signal recovery scheme is $\mathcal{O}(2N^3+4NMKP+NMT+NM+MT)$.

\section{Performance Evaluation Results}\label{sec:simulation}
In this section, we present computer simulation results for the performance evaluation of the proposed joint channel estimation and signal recovery scheme. We have particularly simulated the Normalized Mean Squared Error (NMSE) of the proposed scheme using the metrics $\|\mathbf{H}^{r}-\widehat{\mathbf{H}}^{r}\|^{2}\|\mathbf{H}^{r}\|^{-2}$, $\|\mathbf{H}^{s}-\widehat{\mathbf{H}}^{s}\|^{2}\|\mathbf{H}^{s}\|^{-2}$, and $\|\mathbf{X}-\widehat{\mathbf{X}}\|^{2}\|\mathbf{X}\|^{-2}$. The scaling ambiguity in our two-stage estimation has been removed by normalizing the first column of $\mathbf{H}^r$, such that it has all one elements. For $\mathbf{\Phi}$, we have considered the discrete Fourier transform matrix, which satisfies $\mathbf{\Phi}^H\mathbf{\Phi}=\mathbf{I}_{N}$ and has been suggested as a good choice for ALS \cite{rong2012channel}. All NMSE curves were obtained after averaging over $500$ independent Monte Carlo channel realizations. We have used $\epsilon=10^{-5}$ and $I_\text{max}=15$ in all NMSE performance curves.

\begin{figure} \vspace{-2mm}
	\begin{center}
		\includegraphics[width=0.44\textwidth]{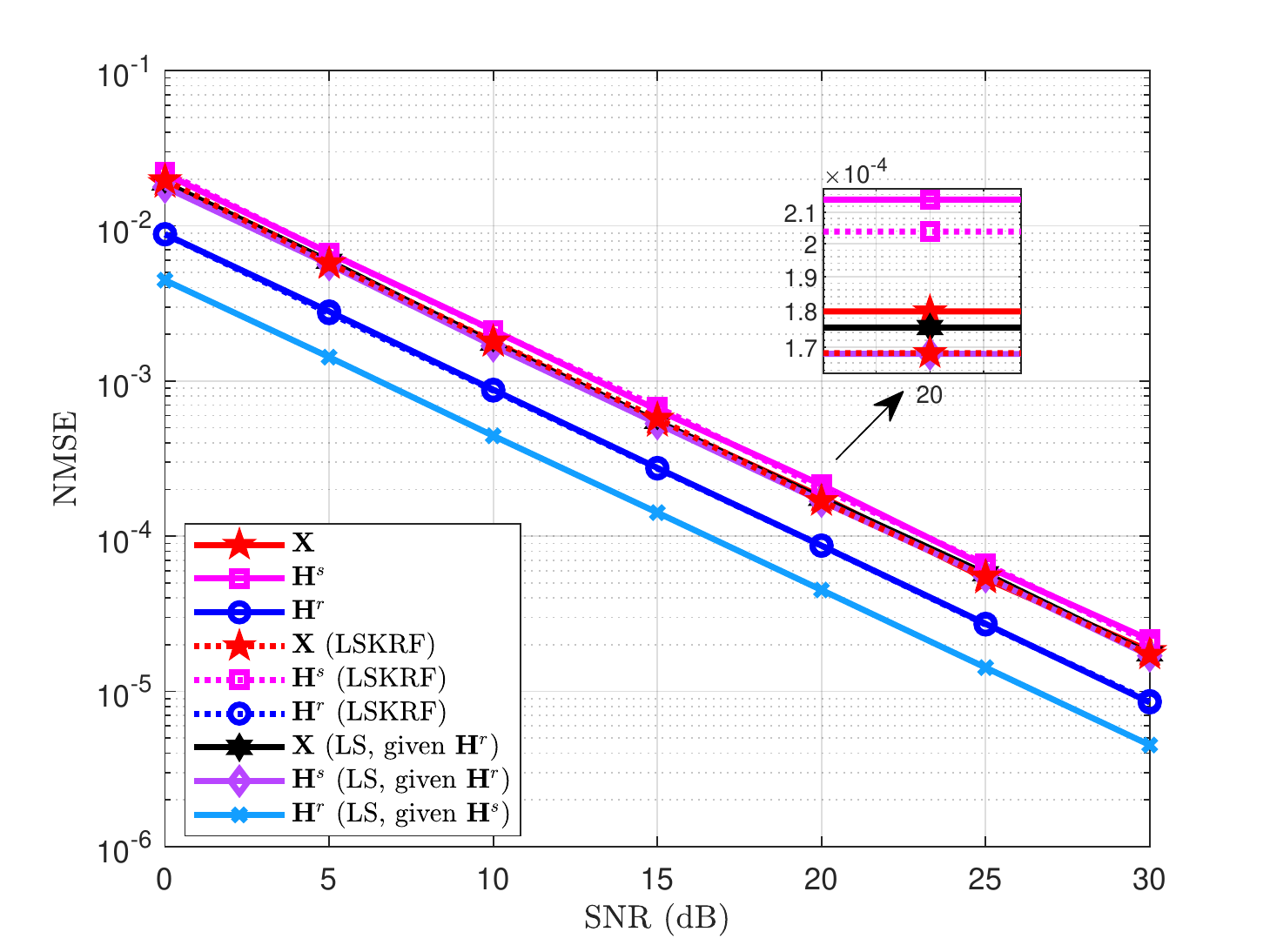}  \vspace{-3mm}
		\caption{NMSE performance comparisons between the proposed algorithm, LSKRF \cite{9104260}, and genie-aided LS estimation versus the SNR in dB for $K=32$, $N=16$, $M=12$, $N=16$, $T=100$, $P=16$, and $\beta=0.2$.}
		\label{fig:com_ls1} \vspace{-6mm}
	\end{center}
\end{figure}

\begin{figure}\vspace{-2mm}
	\begin{center}
		\centerline{\includegraphics[width=0.44\textwidth]{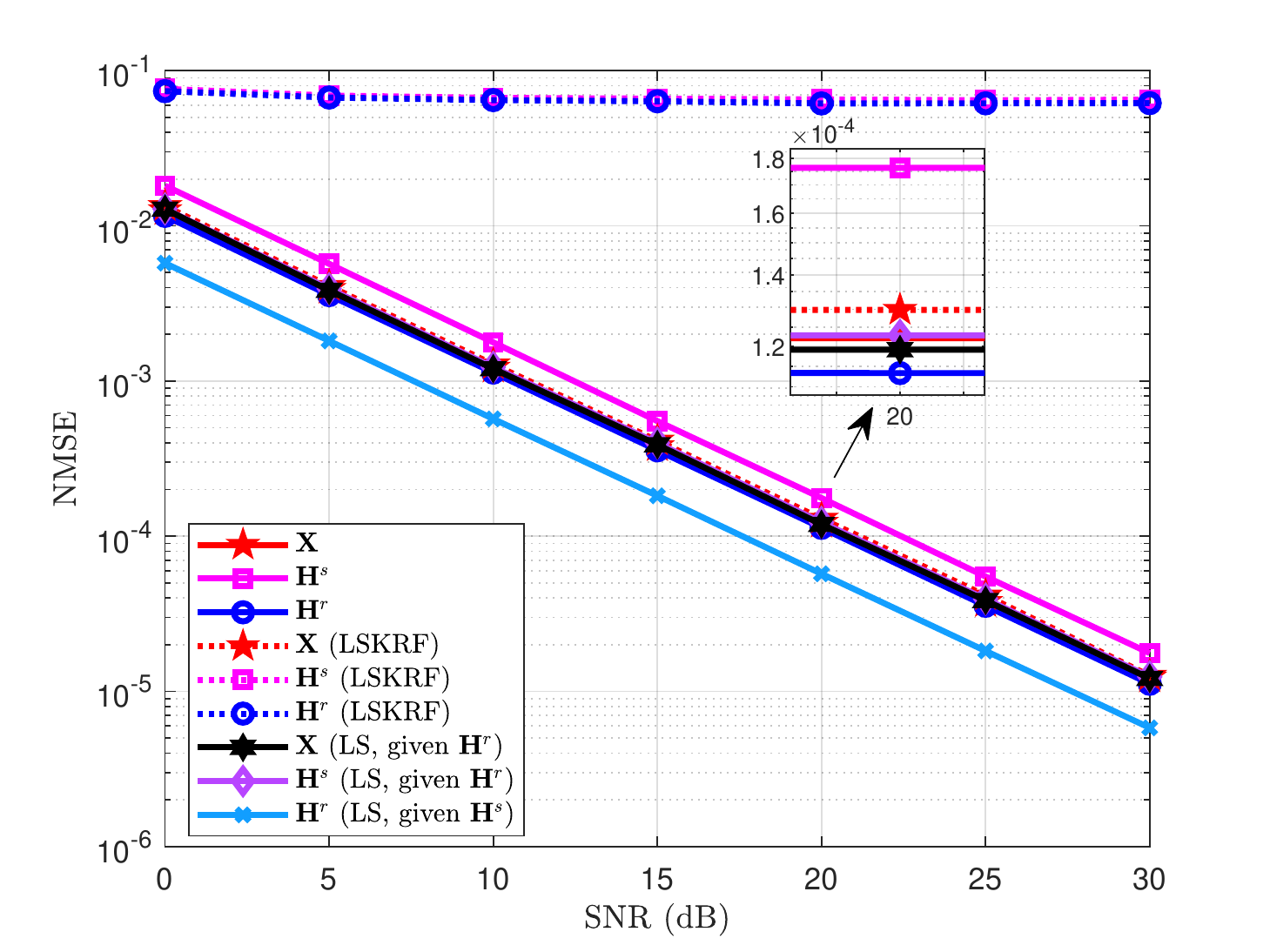} } \vspace{-3mm}
		\caption{NMSE performance comparisons between the proposed algorithm, LSKRF \cite{9104260}, and genie-aided LS estimation versus the SNR in dB for $K=32$, $N=20$, $M=12$, $N=16$, $T=100$, $P=16$, and $\beta=0.2$.}
		\label{fig:com_ls2} \vspace{-6mm}
	\end{center}
\end{figure}

The performance evaluation of the proposed scheme versus the Signal-to-Noise Ratio (SNR) is given in Figs$.$~\ref{fig:com_ls1} and~\ref{fig:com_ls2}. Considering that the joint estimation and recovery in RIS-assisted system has not been studied before, we adopt the channel estimation methods: \textit{i}) LS Khatri-Rao factorization (LSKRF) \cite{9104260}; and \textit{ii}) conventional LS estimation, for the estimation part in the benchmark schemes. In Fig$.$~\ref{fig:com_ls1}, using the setting $K=32$, $N=16$, $M=12$, $N=16$, $T=100$, $P=16$, and $\beta=0.2$, it is shown that proposed scheme achieves similar performance with the one that adopts LSKRF estimation. However, our scheme is more robust to parameter setting changes compared to the one with LSKRF estimation. As depicted in Fig$.$~\ref{fig:com_ls2}, the parameter setting is changed to $K=32$, $N=20$, $M=12$, $N=16$, $T=100$, $P=16$, and $\beta=0.2$. It is observed that the estimation performance of the LSKRF scheme gets dramatically deteriorated with slight changes of $N$, witnessing its increased sensitivity in the parameter setting. More details for solely the channel estimation comparison between our scheme and LSKRF can be found in \cite{wei2020channel}. In addition, the illustrated abnormality in the estimation of the signal $\mathbf{X}$ happens due to its high sparsity and pilots. It is shown in Figs$.$~\ref{fig:com_ls1} and~\ref{fig:com_ls2} that the proposed scheme has only about $2.5$dB performance gap from the LS scheme that assumes perfect channel knowledge. This behavior substantiates the robustness and favorable performance of our proposed scheme.

\begin{figure} \vspace{-2mm}
	\begin{center}
		\includegraphics[width=0.44\textwidth]{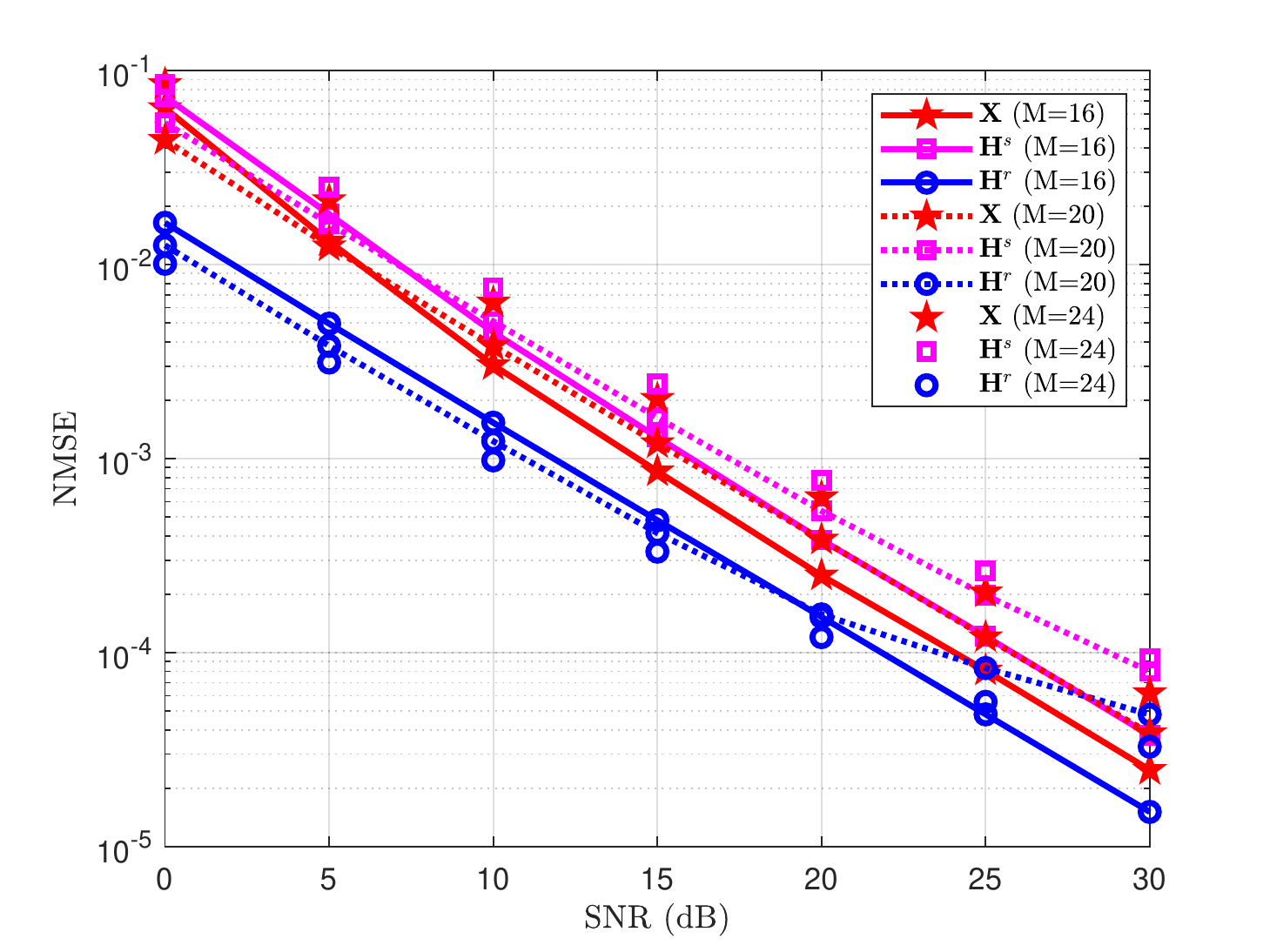}  \vspace{-3mm}
		\caption{NMSE performance of the proposed algorithm versus the SNR in dB for $K=16$, $N=32$, $T=100$, $P=16$, $\beta=0.2$, and various values $M$ for the BS antennas.}
		\label{fig:com_m} \vspace{-6mm}
	\end{center}
\end{figure}

\begin{figure} \vspace{-2mm}
	\begin{center}
		\includegraphics[width=0.44\textwidth]{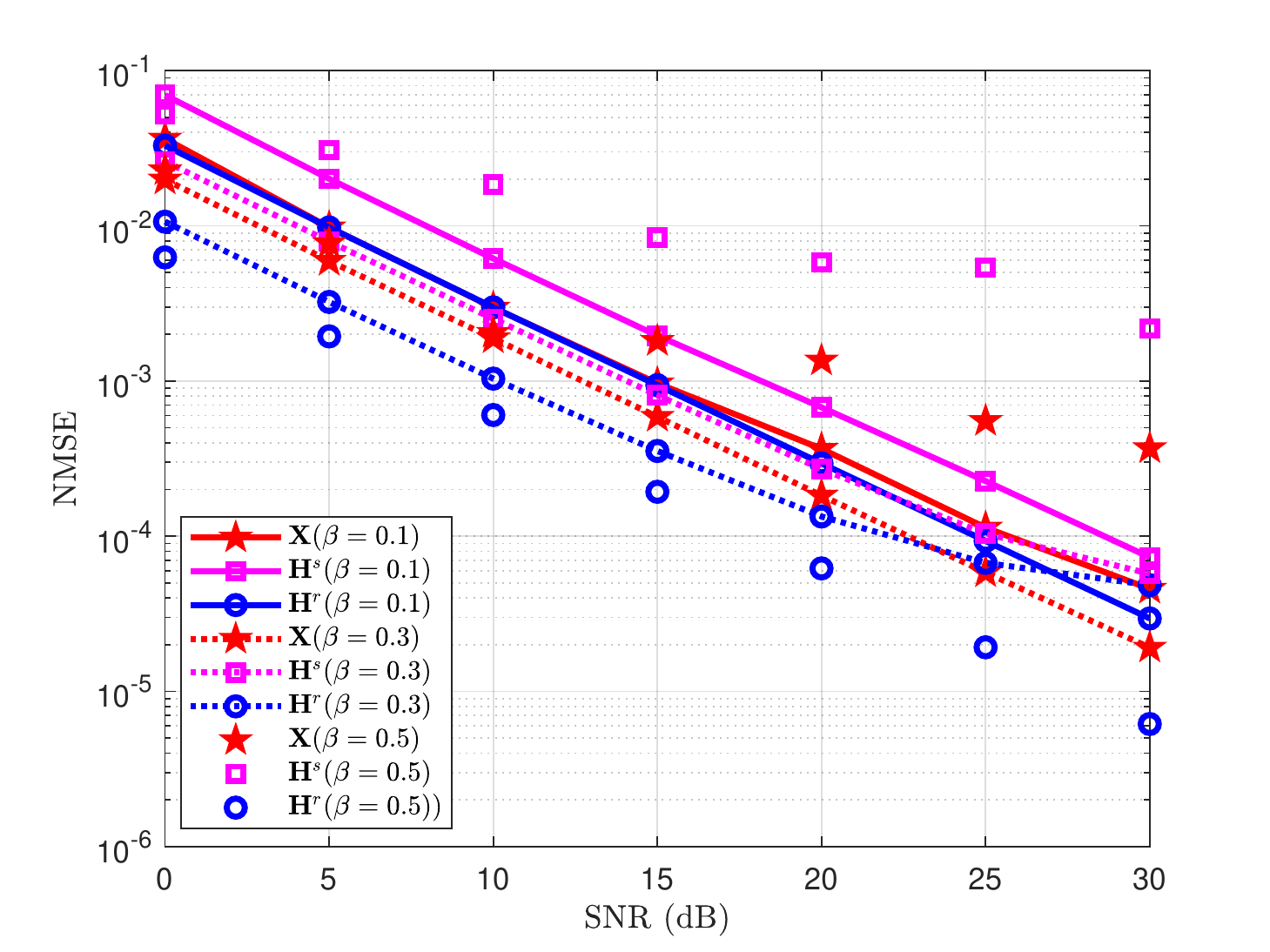}  \vspace{-3mm}
		\caption{NMSE performance of the proposed algorithm versus the SNR in dB for $K=20$, $N=32$, $M=16$, $T=100$, $P=16$, and various values $\beta$ for the sparsity parameter.}
		\label{fig:com_beta} \vspace{-6mm}
	\end{center}
\end{figure}

In Fig$.$~\ref{fig:com_m}, we have used the simulation parameters $K=16$, $N=32$, $T=100$, $P=16$, $\beta=0.2$, and various values of $M$ to illustrate the NMSE performance of the proposed scheme as a function of the SNR. It is evident that there exists an increasing performance loss when $M$ increases. In those cases, the numbers of unknown variables for estimation increase, which result in performance loss.
The impact of the sparsity level $\beta$ in the NMSE performance of the proposed scheme is investigated in Fig$.$~\ref{fig:com_beta} as a function of the SNR for $K=20$, $N=32$, $M=16$, $T=100$, and $P=16$. It is shown that there exists a tradeoff between the NMSE performance and $\beta$. The scheme for $\beta=0.5$ suffers the worst performance. This happens because more non-zero variables need to be estimated with increasing $\beta$, which makes the BiG-AMP algorithm to performs worse. It can be also seen that the performance for $\beta=0.1$ is worse than for $\beta=0.3$ (of around $3$dB), even that less unknown variables are involved. This behavior accounts for the possible sparsity pattern collision. Actually, when multiple users are located close to each other, recovery failures happen \cite{8063440}.

\section{Conclusion}\label{sec:conclusion}
In this paper, we presented a joint channel estimation and signal recovery scheme for RIS-assisted multi-user MISO communication systems, which capitalizes on the PARAFAC decomposition of the received signal model. All involved channels are estimated and the transmitted signal is recovered using a two-stage approach. In the first stage, the channel between RIS and the users and another equivalent channel are estimated using ALS. Then, in the second stage using the equivalent channel estimation, the channel between the BS and RIS and the transmitted signals are recovered via the BiG-AMP algorithm. Our simulation results showcased that the proposed joint channel estimation and signal recovery scheme outperforms various benchmarks. In addition, we observed that the number of BS antennas $M$ and the sparsity level $\beta$ exert a significant effect on our proposed algorithm.

% \newpage

\bibliographystyle{IEEEtran}
\bibliography{strings}

\begin{thebibliography}{10}
\providecommand{\url}[1]{#1}
\csname url@samestyle\endcsname
\providecommand{\newblock}{\relax}
\providecommand{\bibinfo}[2]{#2}
\providecommand{\BIBentrySTDinterwordspacing}{\spaceskip=0pt\relax}
\providecommand{\BIBentryALTinterwordstretchfactor}{4}
\providecommand{\BIBentryALTinterwordspacing}{\spaceskip=\fontdimen2\font plus
\BIBentryALTinterwordstretchfactor\fontdimen3\font minus
  \fontdimen4\font\relax}
\providecommand{\BIBforeignlanguage}[2]{{%
\expandafter\ifx\csname l@#1\endcsname\relax
\typeout{** WARNING: IEEEtran.bst: No hyphenation pattern has been}%
\typeout{** loaded for the language `#1'. Using the pattern for}%
\typeout{** the default language instead.}%
\else
\language=\csname l@#1\endcsname
\fi
#2}}
\providecommand{\BIBdecl}{\relax}
\BIBdecl

\bibitem{Akyildiz2018mag}
I.~F. {Akyildiz}, C.~{Han}, and S.~{Nie}, ``Combating the distance problem in
  the millimeter wave and terahertz frequency bands,'' \emph{IEEE Commun.
  Mag.}, vol.~56, no.~6, pp. 102--108, Jun. 2018.

\bibitem{hu2018beyond}
S.~Hu, F.~Rusek, and O.~Edfors, ``Beyond massive {MIMO}: The potential of
  positioning with large intelligent surfaces,'' \emph{IEEE Trans. Signal
  Process.}, vol.~66, no.~7, pp. 1761--1774, 2018.

\bibitem{huang2019reconfigurable}
C.~Huang, A.~Zappone, G.~C. Alexandropoulos, M.~Debbah, and C.~Yuen,
  ``Reconfigurable intelligent surfaces for energy efficiency in wireless
  communication,'' \emph{IEEE Trans. Wirel. Commun.}, vol.~18, no.~8, pp.
  4157--4170, 2019.

\bibitem{Marco2019}
M.~D. Renzo, M.~Debbah, D.-T. Phan-Huy, A.~Zappone, M.-S. Alouini, C.~Yuen,
  V.~Sciancalepore, G.~C. Alexandropoulos, J.~Hoydis, H.~Gacanin, J.~de~Rosny,
  A.~Bounceur, G.~Lerosey, and M.~Fink, ``Smart radio environments empowered by
  reconfigurable {AI} meta-surfaces: An idea whose time has come,''
  \emph{EURASIP J. Wirel. Commun. Netw.}, vol. 2019, no.~1, pp. 1--20, 2019.

\bibitem{qingqing2019towards}
Q.~Wu and R.~Zhang, ``Towards smart and reconfigurable environment: Intelligent
  reflecting surface aided wireless network,'' \emph{IEEE Commun. Mag.,
  accepted for publication}, 2019.

\bibitem{hcwjasc2020}
C.~{Huang}, R.~{Mo}, and C.~{Yuen}, ``Reconfigurable intelligent surface
  assisted multiuser {MISO} systems exploiting deep reinforcement learning,''
  \emph{IEEE J. Sel. Areas Commun.}, vol.~38, no.~8, pp. 1839--1850, 2020.

\bibitem{alexandropoulos2020reconfigurable}
G.~C. Alexandropoulos, G.~Lerosey, M.~Debbah, and M.~Fink, ``Reconfigurable
  intelligent surfaces and metamaterials: The potential of wave propagation
  control for {6G} wireless communications,'' \emph{IEEE ComSoc TCCN
  Newsletter}, vol.~6, no.~1, June 2020.

\bibitem{9136592}
C.~{Huang}, S.~{Hu}, G.~C. {Alexandropoulos}, A.~{Zappone}, C.~{Yuen},
  R.~{Zhang}, M.~D. {Renzo}, and M.~{Debbah}, ``Holographic {MIMO} surfaces for
  {6G} wireless networks: Opportunities, challenges, and trends,'' \emph{IEEE
  Wirel. Commun.}, vol.~27, no.~5, pp. 118--125, 2020.

\bibitem{9149091}
W.~{Yan}, X.~{Yuan}, Z.~{He}, and X.~{Kuai}, ``Large intelligent surface aided
  multiuser {MIMO}: Passive beamforming and information transfer,'' in
  \emph{2020 ICC}, 2020, pp. 1--7.

\bibitem{tang2019wireless}
W.~Tang, M.~Chen, X.~Chen, J.~Dai, Y.~Han, M.~D. Renzo, Y.~Zeng, S.~Jin,
  Q.~Cheng, and T.~Cui, ``Wireless communications with reconfigurable
  intelligent surface: Path loss modeling and experimental measurement,''
  \emph{[online] https://arxiv.org/abs/1911.05326}, 2019.

\bibitem{huang2018achievable}
C.~Huang, A.~Zappone, M.~Debbah, and C.~Yuen, ``Achievable rate maximization by
  passive intelligent mirrors,'' in \emph{IEEE ICASSP}.\hskip 1em plus 0.5em
  minus 0.4em\relax IEEE, 2018, pp. 3714--3718.

\bibitem{husha_LIS2}
S.~{Hu}, F.~{Rusek}, and O.~{Edfors}, ``Beyond massive {MIMO}: The potential of
  data-transmission with large intelligent surfaces,'' \emph{IEEE Trans. Signal
  Process.}, vol.~66, no.~10, pp. 2746--2758, May 2018.

\bibitem{huang2018energy}
C.~{Huang}, G.~C. {Alexandropoulos}, A.~{Zappone}, M.~{Debbah}, and C.~{Yuen},
  ``Energy efficient multi-user {MISO} communication using low resolution large
  intelligent surfaces,'' in \emph{2018 IEEE GC Wkshps}, Dec 2018, pp. 1--6.

\bibitem{han2019}
Y.~Han, W.~Tang, S.~Jin, C.~Wen, and X.~Ma, ``Large intelligent
  surface-assisted wireless communication exploiting statistical {CSI},''
  \emph{IEEE Trans. Veh. Tech.}, vol.~68, no.~8, pp. 8238--8242, Aug 2019.

\bibitem{yan2019passive}
W.~Yan, X.~Kuai, X.~Yuan \emph{et~al.}, ``Passive beamforming and information
  transfer via large intelligent surface,'' \emph{[online]
  https://arxiv.org/abs/1905.01491}, 2019.

\bibitem{wang2019intelligent}
P.~Wang, J.~Fang, X.~Yuan, Z.~Chen, H.~Duan, and H.~Li, ``Intelligent
  reflecting surface-assisted millimeter wave communications: Joint active and
  passive precoding design,'' \emph{[online] https://arxiv.org/abs/1908.10734},
  2019.

\bibitem{ning2019intelligent}
B.~Ning, Z.~Chen, W.~Chen, and J.~Fang, ``Intelligent reflecting surface design
  for {MIMO} system by maximizing sum-path-gains,'' \emph{[online]
  https://arxiv.org/abs/1909.07282}, 2019.

\bibitem{Alkhateeb2019}
A.~Taha, M.~Alrabeiah, and A.~Alkhateeb, ``Enabling large intelligent surfaces
  with compressive sensing and deep learning,'' \emph{[online]
  https://arxiv.org/abs/1904.10136}, 2019.

\bibitem{huang2019spawc}
C.~{Huang}, G.~C. {Alexandropoulos}, C.~{Yuen}, and M.~{Debbah}, ``Indoor
  signal focusing with deep learning designed reconfigurable intelligent
  surfaces,'' in \emph{IEEE 20th SPAWC}, July 2019, pp. 1--5.

\bibitem{9053976}
G.~C. {Alexandropoulos} and E.~{Vlachos}, ``A hardware architecture for
  reconfigurable intelligent surfaces with minimal active elements for explicit
  channel estimation,'' in \emph{IEEE ICASSP}, Barcelona, Spain, 4–8 May
  2020, pp. 9175--9179.

\bibitem{wei2020channel}
L.~Wei, C.~Huang, G.~C. Alexandropoulos, C.~Yuen, Z.~Zhang, M.~Debbah
  \emph{et~al.}, ``Channel estimation for {RIS}-empowered multi-user {MISO}
  wireless communications,'' \emph{arXiv preprint arXiv:2008.01459}, 2020.

\bibitem{9104260}
G.~T. {de Araújo} and A.~L.~F. {de Almeida}, ``{PARAFAC}-based channel
  estimation for intelligent reflective surface assisted {MIMO} system,'' in
  \emph{2020 IEEE 11th SAM}, 2020, pp. 1--5.

\bibitem{6898015}
J.~T. {Parker}, P.~{Schniter}, and V.~{Cevher}, ``Bilinear generalized
  approximate message passing—part i: Derivation,'' \emph{IEEE Trans. Signal
  Processing}, vol.~62, no.~22, pp. 5839--5853, 2014.

\bibitem{8580585}
X.~{Meng} and J.~{Zhu}, ``Bilinear adaptive generalized vector approximate
  message passing,'' \emph{IEEE Access}, vol.~7, pp. 4807--4815, 2019.

\bibitem{8357527}
------, ``A generalized sparse bayesian learning algorithm for 1-bit doa
  estimation,'' \emph{IEEE Commun. Lett.}, vol.~22, no.~7, pp. 1414--1417,
  2018.

\bibitem{7574287}
S.~{Wu}, L.~{Kuang}, Z.~{Ni}, D.~{Huang}, Q.~{Guo}, and J.~{Lu},
  ``Message-passing receiver for joint channel estimation and decoding in {3D}
  massive { MIMO-OFDM} systems,'' \emph{IEEE Trans. Wirel. Commun.}, vol.~15,
  no.~12, pp. 8122--8138, 2016.

\bibitem{sidiropoulos2000blind}
N.~D. Sidiropoulos, G.~B. Giannakis, and R.~Bro, ``Blind {PARAFAC} receivers
  for {DS-CDMA} systems,'' \emph{IEEE Trans. Signal Process.}, vol.~48, no.~3,
  pp. 810--823, 2000.

\bibitem{kolda2009tensor}
T.~G. Kolda and B.~W. Bader, ``Tensor decompositions and applications,''
  \emph{SIAM review}, vol.~51, no.~3, pp. 455--500, 2009.

\bibitem{kroonenberg1980principal}
P.~M. Kroonenberg and J.~De~Leeuw, ``Principal component analysis of three-mode
  data by means of alternating least squares algorithms,''
  \emph{Psychometrika}, vol.~45, no.~1, pp. 69--97, 1980.

\bibitem{rong2012channel}
Y.~Rong, M.~R. Khandaker, and Y.~Xiang, ``Channel estimation of dual-hop {MIMO}
  relay system via parallel factor analysis,'' \emph{IEEE Trans. Wirel.
  Commun.}, vol.~11, no.~6, pp. 2224--2233, 2012.

\bibitem{harshman1994parafac}
R.~A. Harshman and M.~E. Lundy, ``{PARAFAC}: Parallel factor analysis,''
  \emph{Computational Statistics \& Data Analysis}, vol.~18, no.~1, pp. 39--72,
  1994.

\bibitem{bro2003new}
R.~Bro and H.~A. Kiers, ``A new efficient method for determining the number of
  components in {PARAFAC} models,'' \emph{J. Chemom.: A Journal of the
  Chemometrics Society}, vol.~17, no.~5, pp. 274--286, 2003.

\bibitem{ten2002uniqueness}
J.~M. ten Berge and N.~D. Sidiropoulos, ``On uniqueness in
  {CANDECOMP/PARAFAC},'' \emph{Psychometrika}, vol.~67, no.~3, pp. 399--409,
  2002.

\bibitem{roemer2008closed}
F.~Roemer and M.~Haardt, ``A closed-form solution for parallel factor
  ({PARAFAC}) analysis,'' in \emph{IEEE ICASSP}.\hskip 1em plus 0.5em minus
  0.4em\relax IEEE, 2008, pp. 2365--2368.

\bibitem{lioliou2009performance}
P.~{Lioliou}, M.~{Viberg}, and M.~{Coldrey}, ``Performance analysis of relay
  channel estimation,'' in \emph{2009 Conference Record of the Forty-Third
  ACSSC}, Nov 2009, pp. 1533--1537.

\bibitem{8879620}
Z.~{He} and X.~{Yuan}, ``Cascaded channel estimation for large intelligent
  metasurface assisted massive {MIMO},'' \emph{IEEE Wirel. Commun. Lett.},
  vol.~9, no.~2, pp. 210--214, 2020.

\bibitem{8063440}
J.~{Zhang}, X.~{Yuan}, and Y.~A. {Zhang}, ``Blind signal detection in massive
  {MIMO}: Exploiting the channel sparsity,'' \emph{IEEE Trans. Commun.},
  vol.~66, no.~2, pp. 700--712, 2018.

\end{thebibliography}
\vspace{12pt}

\end{document}